\input{aipcheck}

\documentclass[
    ,final            % use final for the camera ready runs
  ]
  {aipproc}

\layoutstyle{8x11double}

\newcommand{\etc}{{\rm ETC}}
\def\gsim{\mathrel{\raise.3ex\hbox{$>$\kern-.75em\lower1ex\hbox{$\sim$}}}}

\begin{document}

\title[Lattice Technicolor]{QCD-like technicolor on the lattice}


\author{Kari Rummukainen}{
  address={Department of Physics and Helsinki Institute of Physics, 
    P.O.Box 64, 00014 University of Helsinki,
    Finland}
}

%\author{<author2>}{
%  address={<common address for author2 and author3>}
%}

%\author{<author3>}{
%  address={<common address for author2 and author3>}
%  ,altaddress={<author1 address>} % additional visiting address
%}

\begin{abstract}
  This talk gives an overview, aimed at non-experts,
  of the recent progress of technicolor models on the lattice.
  Phenomenologically successful technicolor models require 
  walking coupling; thus, an emphasis is put on the determination
  of the $\beta$-function of various models.
  As a case study we consider SU(2) gauge field theory with
  two adjoint representation fermions, so-called minimal
  walking technicolor theory.
\end{abstract}

\maketitle

%%%%%%%%%%%%%%%%%%%%%%%%%%%%%%%%%%%%%%%%%%%
%% MAINMATTER
%%%%%%%%%%%%%%%%%%%%%%%%%%%%%%%%%%%%%%%%%%%%

\section{Walking technicolor}

The Standard Model, with the Higgs field, is phenomenologically
extremely successful.  However, the Higgs field has a very special
status in the Standard Model (SM): it provides the mechanism for the
electorweak symmetry breaking, which is a central feature of the
Standard Model.  It also is the sole scalar field of the theory, which
leads to well-known theoretical problems at energies much higher than
the electroweak scale: the hierarchy problem, the stability of the
electroweak vacuum, triviality bounds etc.  Indeed, many extensions of
the SM are motivated by the amelioration of these problems.

In technicolor \cite{Weinberg,Susskind} the symmetry breaking with
the fundamental Higgs scalar is substituted with a QCD-like chiral
condensate.  We remind that QCD already contributes to the electroweak
symmetry breaking: the chiral $\bar q q$ condensate has electroweak
charge, which, if we would remove the Higgs field from the standard
model, would alone break the electroweak gauge symmetry and give rise
to $W$ and $Z$ boson masses of order $f_{\pi}$ (and 3 $\pi$-mesons
would be absorbed as the longitudinal polarizations of $W$, $Z$).
In technicolor this mechanism is transferred  to the electroweak scale
$\Lambda_{\rm EW}$.  Thus, we introduce a new non-Abelian gauge field,
{\em technigauge}, and massless fermions, {\em techniquarks} $Q$.
Techiquarks are assumed to have both technicolor and electroweak
charge, exactly like quarks in the Standard Model, and they form
chiral $\bar Q Q$ condensate which breaks the electroweak symmetry.
The magnitude of the chiral condensate (or rather, the magnitude of
the decay constant $f_{\rm TC}$) takes the role of the Higgs
condensate ($\sim 256\,$GeV).  Three of the would-be Goldstone bosons,
pseudoscalar technimesons, become the longitudinal polarizations of
the $W$ and $Z$ bosons.  As with QCD, naturally also technicolor has
several bound states which would be observable in experiments. This is
an elegant and proven mechanism which, in a natural way, describes the
electroweak gauge and Higgs sector without a fundamental scalar
particle.

However, the above classic technicolor scenario does not provide for
the Standard Model fermion mass terms.  This is addressed in the {\em
  extended technicolor} theories \cite{Eichten:1979ah,Hill:2002ap},
which produce a Yukawa-like coupling to the technifermion
condensate. This can be modeled with a gauge boson with mass $M_\etc$,
and which is coupled to the SM fermions (denoted by $q$) and
techniquarks ($Q$):

\centerline{ \includegraphics[width=0.45\columnwidth]{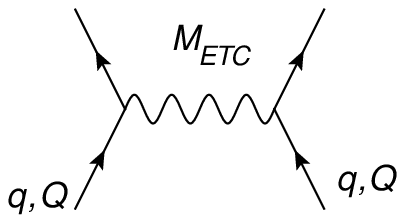} }

At energies smaller than $M_\etc$, this leads to effective four-fermi
couplings:
\begin{itemize}
\item
  $g_\etc^2/M_\etc^2 \bar QQ \bar qq$, which gives
  fermion masses 
  $m_q \propto \langle \bar Q Q\rangle/M_\etc^2$.
\item
  $g_\etc^2/M_\etc^2 \bar qq \bar qq$, which contributes
  to unwanted flavour changing neutral currents.
%\item
%  $g_\etc^2/M_\etc^2 \bar QQ \bar QQ$, which
%  causes explicit chiral symmetry breaking in the techniquark
%  sector
\end{itemize}
Precision electroweak measurements strongly constrain flavour changing
neutral currents, and this gives the generic constraint $M_\etc \gsim
1000\Lambda_{\rm EW}$.  We recall that the electroweak symmetry
breaking pattern requires $\langle \bar Q Q\rangle_{\rm TC} \propto
\Lambda_{\rm TC}^3 \approx \Lambda_{\rm EW}^3$, where
$\langle\rangle_{\rm TC}$ indicates that the condensate is evaluated
at the technicolor ($\sim$ electroweak) scale.  Naively, the above
conditions will give too small SM fermion masses.  This can be
circumvented if the condensate at the extended technicolor scale is
enhanced: we need $\langle \bar Q Q\rangle_{\rm ETC} \propto m_q
\Lambda_{\rm ETC}^2$.

The renormalization group evolution of the technifermion condensate 
is
\begin{equation}
  \langle \bar Q Q \rangle_{\rm ETC} = 
  \langle \bar Q Q\rangle_{\rm TC} 
  \exp\left[\int_{\Lambda_{\rm TC}}^{M_{\rm ETC}} 
    \frac{\gamma(g^2)}{\mu} d\mu\right]
\end{equation}
In a weakly coupled theory the anomalous exponent $\gamma \sim 0$, and
the condensate $\langle \bar Q Q\rangle$ remains approximately
constant.  Thus, it is not possible to satisfy the constraints in a
QCD-like theory, where the coupling remains large only in a narrow
energy range above the chiral symmetry breaking.

\begin{figure}
  \hfill\includegraphics[width=0.9\columnwidth]{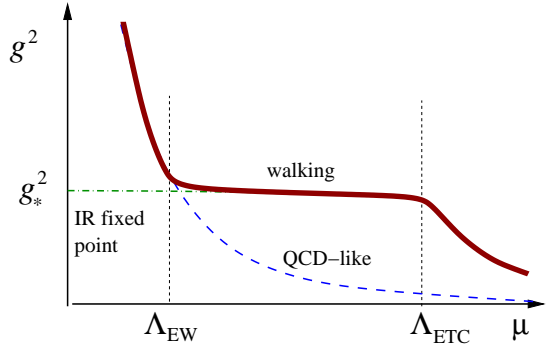}
  \caption{The evolution of the walking coupling, versus
    the QCD-like coupling (dashed line)
    and the coupling with an infrared
    fixed point (dash-dotted).}
  \label{fig:coupling}
\end{figure}

However, if the coupling constant {\em walks} 
\cite{Holdom:1981rm,Yamawaki:1985zg,Appelquist:an}, i.e.  $g^2$ remains
almost constant $\sim g_*^2$ over the whole energy range from TC to
ETC, then
\begin{equation}
  \langle \bar Q Q\rangle_{\rm ETC} \approx
    \left(\frac{\Lambda_{\rm ETC}}{\Lambda_{\rm TC}}\right)^{\gamma(g^2_*)}
    \langle \bar Q Q\rangle_{\rm TC} 
    \label{evolution}
\end{equation}
This is called condensate enhancement.  The desired behaviour of the
coupling is shown in Figure~\ref{fig:coupling}. This makes it possible
to satisfy the constraints above, if the anomalous exponent
$\gamma(g^2_*)$ is large enough, 1--2.

\begin{figure}
  \centerline{\includegraphics[width=0.9\columnwidth]{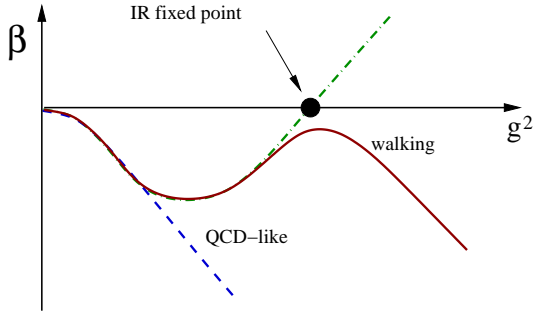}}
  \caption{The schematic $\beta$-functions of theories leading
    to behaviour shown in Figure~\ref{fig:coupling}.}
  \label{fig:betaf}
\end{figure}

The $\beta$-function
\begin{equation}
  \beta = \mu \frac{{\rm d}g}{{\rm d}\mu}
\end{equation}
of the walking theory is shown in Figure~\ref{fig:betaf}.  The
$\beta$-function almost reaches zero at some value of the coupling.
If the $\beta$-function reaches zero at $g^2=g_*^2$, the theory has an
infrared fixed point (IRFP) and the long distance behaviour is
conformal, scale invariant.  We remind that the value of the coupling
at the fixed point is regularisation scheme dependent, but its
existence is not.  A theory with an IRFP can be deformed to the
walking case by introducing a scale, e.g by adding an arbitrarily
small mass term for the techniquarks.  Thus, both walking theory or a
theory with an IRFP are suitable starting points for a technicolor model.

What kind of theories exhibit walking behaviour?  Guidance is
given by the 2-loop scheme-invariant $\beta$-function of SU($N$) gauge
theory with $N_f$ fermions of representation $R$:
\begin{equation}
  \beta(g)=-\mu\frac{d g}{d\mu} =
  -\beta_0\frac{g^3}{16\pi^2}-\beta_1\frac{g^5}{(16\pi^2)^2}
  \label{pertbeta}
\end{equation}
where $\beta_0 =\frac{11}{3}N-\frac{4}{3}T(R) N_f$, $\beta_1
=\frac{34}{3}N^2 -\frac{20}{3} N T(R) N_f - 4C_2(R) T(R) N_f$,
 $C_2(R)$ is the second Casimir invariant of the fermion
representation $R$, and
$T(R)\delta^{ab} = {\rm Tr\,}T^aT^b$.
In asymptotically free theories $\beta_0$ must be positive.
if $\beta_1 < 0$, the theory is QCD-like; and if $\beta_1 >0$,
then Eq.~(\ref{pertbeta}) gives an IRFP at some coupling
(Banks-Zaks fixed point \cite{Banks:1981nn}).
However, generically this fixed point is at strong
coupling, and the perturbative analysis is not valid.  It is
expected that at strong coupling the theory has chiral symmetry
breaking, and therefore,
the actual {\em conformal window} where the theory has IRFP is
expected to be narrower.
 
\begin{figure}
\hfill\includegraphics[width=\columnwidth]{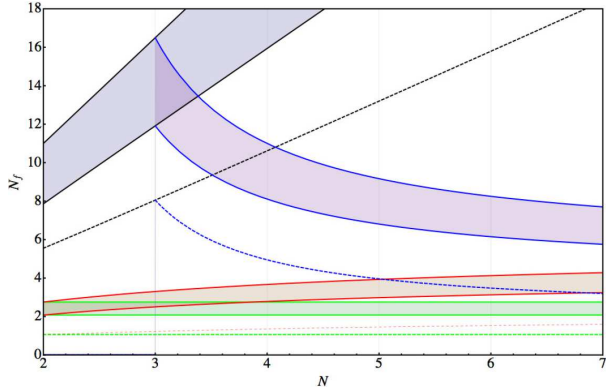}\hfill
  \caption{The conformal window in SU($N$) gauge field theory
    with $N_f$ fermion flavours in different fermion representations.
    From top to bottom, the bands correspond to the fundamental,
    two-index antisymmetric, two-index symmetric and to the adjoint
    represenation fermions.
    The upper edge of the bands correspond to the loss of the
    asymptotic freedom, and the lower edge of the band
    has been calculated using the ladder approximation
    \cite{Sannino:2004qp}.
}
  \label{fig:confwin}
\end{figure}

The situation in SU($N$) gauge theories with $N_f$ flavours of
fundamental, 2-index antisymmetric, 2-index symmetric and adjoint
representation fermions is summarized in Figure~\ref{fig:confwin}.
The conformal windows are shown with shaded regions: the upper edges
are where $\beta_0$ changes sign, and the lower edges have been
estimated with ladder approximation \cite{Sannino:2004qp}; below
this line the system is expected to have chiral symmetry breaking.
The lines below the shaded regions show the point where $\beta_2$ changes
sign.  Thus, walking coupling can be expected to be found near the lower
edge of the conformal window.

It turns out that the best fit to 
electroweak precision measurements is obtained 
using higher than fundamental representations, because in this
case the conformal window is reached with a smaller number of fermions
\cite{Sannino:2004qp}.
This means that the adjoint representation and
2-index symmetric representation  are the most interesting;
indeed, the two favourite theories are
SU(2) gauge with two adjoint fermions (``minimal walking technicolor'', MWTC),
and SU(3) with two 2-index symmetric representation fermions.

\section{Theories studied on the lattice}

If a theory is to be a candidate for walking technicolor,
it should be in close proximity of the lower edge
of the conformal window.
It should be either just below the window, in which case 
it exhibits the walking coupling behaviour, 
or just within the conformal window.
In the latter case the theory is easy to deform into a walking 
theory by adding a mass or momentum scale to it, e.g. an explicit 
arbitrarily small mass term for the technifermions.  At energy scales
less than the mass term the physics is dominated by the gauge fields
and the theory confines.
Thus, one of the main goals of lattice simulations is to measure
the $\beta$-function of the theory. 
% The anomalous 
%exponent of the mass is also important for technicolor.  

Of the theories shown in Figure~\ref{fig:confwin} the
ones with fermions in the fundamental representation are
the most familiar, due to their relation with QCD.
The most studied case is SU(3) gauge theory with $N_f=8$--$16$ flavours
of fundamental fermions
\cite{Damgaard:1997ut,
  Appelquist:2007hu, Appelquist:2009ty,Fodor:2009wk,Fodor:2010zz,
  Deuzeman:2008sc, Deuzeman:2009mh,
  Deuzeman:2010zz, Itou:2010we, Jin:2010vm, Hayakawa:2010yn,Hasenfratz:2010fi}.
These studies indicate that for $N_f \le 9$ the theory has chiral
symmetry breaking and when $N_f \le 10 \le 16$ it has an IRFP
(however, at $N_f =12$ results are conflicting, indicating that
the systematics with different methods are not yet fully under control).
At $N_f>16$ the theory loses asymptotic freedom.

With SU(2) and fundamental representation fermions the
results indicate that for $6 \le N_f \le 10$ the theory is within the
conformal window \cite{Bursa:2010xr,Ohki:2010sr}.  It is not currently
known where the lower edge of the conformal window is.

Most interesting theories for technicolor phenomenology are the ones
which are at the conformal window with smaller number of flavours.
SU(2) with two adjoint representation fermions has been studied by
several groups
\cite{Catterall:2007yx,Hietanen:2008mr,DelDebbio:2008zf,Catterall:2008qk,
  Hietanen:2009az,Bursa:2009we,DelDebbio:2009fd}.  As described in
detail below, there are strong indications that SU(2) with two adjoint
fermions is within the conformal window.

There also exist a number of studies of SU(3) gauge with two 2-index 
symmetric representation fermions \cite{Shamir:2008pb,DeGrand:2010na,
  Fodor:2009ar,Kogut:2010cz}.  In this case the existence of 
an infrared fixed point is not fully clarified.

\section{Minimal walking technicolor}

Let us have a closer look at SU(2) gauge field
theory with two adjoint fermions, ``minimal walking
technicolor'' MWTC, and more precisely, the determination
of the $\beta$-function, anomalous exponent and the mass spectrum. 

\subsection{$\beta$-function in MWTC}

Several methods have been used to measure the $\beta$-function
in the theories discussed here: step
scaling with Schr\"odinger functional scheme
\cite{Appelquist:2007hu,Shamir:2008pb,Hietanen:2009az,
  Appelquist:2009ty,Bursa:2009we,
  DeGrand:2010na,Hayakawa:2010yn}, Monte Carlo
renormalization group \cite{Hasenfratz:2010fi,Catterall:2010du}, 
Twisted polyakov
and twisted Wilson loop schemes \cite{Ohki:2010sr,Itou:2010we}.  All
of these schemes are based on the response of the system when a length
scale, typically lattice size, is changed.  Here we shall
look at the Schr\"odinger functional method.

In the Schr\"odinger functional \cite{Luscher:1992an}
we introduce a constant background field using special
boundary conditions and measure the response of the
system when the background field is changed.
The method was developed and very succesfully applied to
QCD by the Alpha collaboration \cite{Luscher:1992zx,Luscher:1993gh,
DellaMorte:2004bc}.  One of the advantages of this method is that
simulations with exactly massless fermions are possible: the fixed boundary
conditions regulate the eigenvalues of the fermion (Dirac) matrix.
Thus, the often cumbersome extrapolation to vanishing fermion masses
is avoided.  

Consider lattices of volume 
$V=L^4=(Na)^4$, where $a$ is the lattice spacing, 
and following \cite{Luscher:1992zx}, 
the spatial gauge links on the $x_0=0$ and $x_0=L$ are fixed so
so that we obtain color diagonal boundary gauge fields
\begin{eqnarray}
  A_i(x_0 = 0) &=& \eta\sigma_3/(g_0 L) \\
  A_i(x_0 = L) &=& (\pi-\eta)\sigma_3/(g_0 L)
%  U_\mu(x_0 = 0) &=& \exp(-i\eta\sigma_3 a/L)\\
%  U_\mu(x_0 = L) &=& \exp(-i(\pi-\eta)\sigma_3 a/L).
\end{eqnarray}
Here $\sigma_3$ is the third Pauli matrix, and $g_0$ is
the bare gauge coupling.
The spatial gauge field boundary conditions are periodic.
%During the simulation the fermion fields are set to vanish at 
%boundaries $x_0=0$ and $x_0=L$,
%and are ``twisted'' periodic to spatial directions:
%$\psi(x+L\hat e_i) = \exp(i\pi/5)\psi(x)$ \cite{Sint:1995rb}. 

At the classical level the boundary conditions above
generate a constant Abelian chromoelectric background field, which
has an easily calculabe action.  The derivative
of the action with respect to $\eta$ is 
%\begin{equation}
${\partial S^{\textrm{class.}}}/{\partial\eta} =
{k(N,\eta)}{g_{0}^{-2}}$,
%\end{equation}
where $k(N,\eta)$ is a known function.  By generalizing this
to quantum level we obtain the definition for the coupling $g^2(L,a)$:
\begin{equation}
  \left\langle\frac{\partial S}{\partial \eta}\right\rangle =
  \frac{k(N,\eta)}{g^{2}(L)}\,.
 \label{sfcoupling}
\end{equation}
We fix $\eta=\pi/4$ after taking the derivative.
The lhs of Eq.~(\ref{sfcoupling}) gives a boundary operator
which is straightforward to evaluate: in case of unimproved
fermions, it is simply the expectation value of the
plaquettes touching the boundary.  Eq.~(\ref{sfcoupling}) gives the
coupling evaluated at length scale $L$, the lattice size.
Thus, by varying the size of the lattice at fixed lattice
spacing $a$ the evolution of the coupling can be measured.
By using different lattice spacings $a$
(different lattice gauge couplings $g_0$) the physical size
of the lattice can be changed over a very wide range of length
scales.

\begin{figure}
%\hfill\includegraphics[width=0.4\textwidth]{g2_L_su2.eps}
%\hfill\hfill
  \centerline{
    \includegraphics[width=0.85\columnwidth]{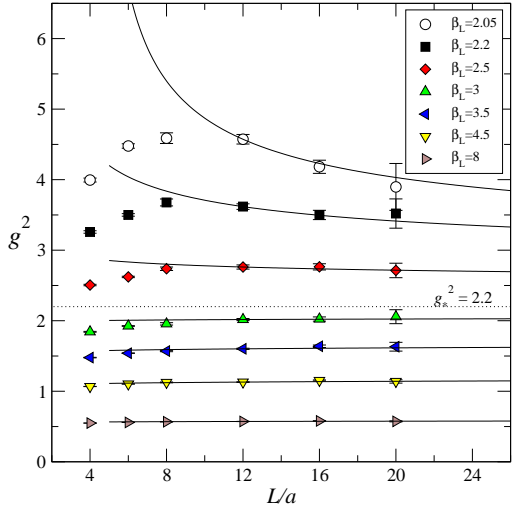}
}
\caption{%{\em Left:} the evolution of the coupling in QCD-like theory,
  %SU(2) with 2 {\em fundamental} fermions.  
  %{\em Right:} 
  The evolution of the measured coupling $g^2$ in MWTC, SU(2) with 
  2 adjoint fermions, in the Scr\"odinger functional scheme.
  Different plot symbols correspond to different bare couplings 
  $\beta_L \equiv 4/g_0^2$ and hence to different lattice spacings $a$.
  Thus, the physical volume between the sets of is very different.
  The continuous lines show the evolution using 
  % 2-loop $\beta$-function (\ref{pertbeta}, {\em left}) or 
  a $\beta$-function
  ansatz, Eq.~(\ref{betansatz}) \cite{Hietanen:2009az}.}
  \label{fig:g2}
\end{figure}

In a QCD-like theory the $\beta$-function is negative, which
means that the coupling constant $g^2(L)$ increases as the 
lattice size is increased while lattice spacing is kept constant. 
This has been observed in simulations of QCD or SU(2) with two
fundamental representation fermions.
%As an example, this can be clearly 
%seen in Figure~\ref{fig:g2}, where the behaviour of 
%$g^2(L)$ in SU(2) gauge theory with two fundamental
%representation fermions is shown.  The behaviour fits
%well with 2-loop perturbative curve.

In contrast, in MWTC the situation is very different, as shown 
in Figure~\ref{fig:g2}.  What is shown here are measurements of $g^2(L)$ 
at seven different lattice spacings $a$  and varying lattice size
$N=L/a=4\ldots 20$.   Thus, at each lattice spacing we obtain
a segment of the evolution when the scale varies over a factor of 
$20/4 = 5$.  Here one observes that when coupling is small
(lower sets of data), $g^2(L)$ increases very slowly 
with increasing lattice size $L/a$.  This is in accord with the 
two-loop perturbative $\beta$-function (\ref{pertbeta}).  However,
when the coupling increases, the growth slows.
Finally, at $g^2 \gsim 3$
the measured coupling $g^2$ {\em decreases} with increasing $L$, if $L$ is
large enough, indicating that the $\beta$-function is positive.
This behaviour is consistent with an infrared fixed 
point at $g_*^2 \approx 2$--$3$.  These results are
in agreement with the results from \cite{Bursa:2009we}.

The lattice spacing is a function of 
the bare lattice gauge coupling $\beta_L \equiv 4/g_0^2$,
but a priori we do not its value.  The measurements
of $g^2(L)$ are used to match the lattices: if, say,
we measure a particular value of $g^2$ on a lattice of size $N$
and coupling $g_0$, and the same value of $g^2$ on a lattice of
size $N/2$ and coupling $g_0'$, we know that the lattice spacing
at $g_0'$ is twice the one at $g_0$.

However, this procedure is made more complicated by 
finite lattice spacing artifacts.
The effect of these can be seen from 
Figure~\ref{fig:g2}, where at large $g^2$ 
the data first increases at small $L/a$ and only starts to
decrease at large $L/a$.  The small $L$ behaviour is a finite lattice
spacing artifact, and for reliable results 
a proper continuum limit extrapolation must be used.
The standard way to achieve this is to use the step scaling
function hierarchially at different lattice sizes (for details,
see e.g. \cite{Luscher:1992zx}).  However, because the small
volume data is very questionable, this was not attempted here.

The main reason for the difficulty of using the step scaling
is that the evolution of $g^2(L)$ in MWTC is very slow, much slower
than in QCD.  Then the evolution is easily masked by 
finite lattice artifacts, which are of similar magnitude as in QCD.
Improved actions can be expected to help: smaller cutoff effects
mean that smaller lattice sizes should become usable 
in continuum extrapolations.

\begin{figure}
  \centerline{\includegraphics[width=0.85\columnwidth]{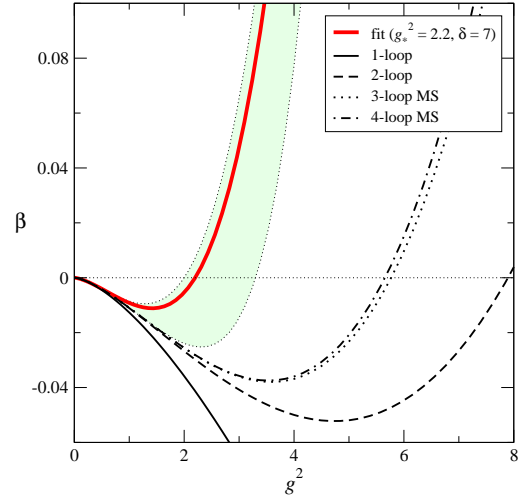}}
  \caption{%
    The $\beta$-function of the MWTC in Schr\"odinger functional scheme
    obtained from the fit to the
    data.  The shaded band shows the estimated error range, and
    the thick line corresponds to fit parameters $g_*^2=2.2$,
    $\delta=7$.  Shown are also the universal 2-loop
    $\beta$-function (dashed line), for comparison,
    3- and 4-loop MS-scheme $\beta$-functions \cite{vanRitbergen:1997va}.
    Figure from ref.~\cite{Hietanen:2009az}.}
  \label{fig:fitbeta}
\end{figure}
Nevertheless, one can use the large volume ($12^4$--$20^4$)
data and check the consistency of the results by fitting
a $\beta$-function ansatz:
\begin{equation}
  \beta = -L\frac{d g}{dL} = -b_1 g^3 - b_2 g^5 - b_3 g^\delta
\label{betansatz}
\end{equation}
Here $b_1$, $b_2$ are perturbative constants and $b_3$ and $\delta$
are fit parameters, parametrising the location of the
fixed point and the slope of the $\beta$-function there.  
Using this fit ansatz the continuous lines in Figure~\ref{fig:g2}
are obtained.  The resulting $\beta$-function is shown in
Figure~\ref{fig:fitbeta}.  The error band of the fitted 
$\beta$-function is very large.

For comparison, in Figure~\ref{fig:fitbeta} also the universal
2-loop $\beta$-function is shown.  It has the Banks-Zaks fixed
point at $g_*^2 \approx 8$, at much stronger coupling than the
result from the simulations, $g_*^2 = 2$--$3$.  The Banks-Zaks
fixed point is at strong coupling, and perturbative analysis is
not reliable.  This is indicated by the 3- and 4-loop MS-scheme
results also shown in figure \cite{vanRitbergen:1997va}; here the
fixed point moves substantially towards smaller coupling.
It must be emphasized that the MS-scheme values are not
directly comparable to the SF-scheme results obtained from the
lattice, but are shown to quantify the perturbative uncertainty.

However, the fact that these results are obtained with unimproved
Wilson fermion action casts doubts about the validity of the
above result: unimproved Wilson action has large $O(a)$ errors,
and, as already described above, these make small lattice sizes
$L/a < 10$ to be unusable.  It would be very welcome to repeat
the analysis using fully $O(a)$ improved action.  Indeed,
large dependence on the action used has been observed in studies
of SU(3) gauge with 2-index symmetri representation
fermions \cite{DeGrand:2010na}.  Steps toward
using non-perturbatively improved action for MWTC have been 
made in ref.~\cite{Karavirta:2010ef}.

A large value for the mass anomalous dimenson 
$\gamma$ around the fixed point
is relevant for the condensate enhancement and thus for the success
of the extended technicolor model.  
This can be measured using related Schr\"odinger functional
method.  For MWTC this has been done by Bursa {\em et al.} \cite{Bursa:2009we},
using the same unimproved Wilson fermion action as was used
in the $\beta$-function study described above.  
They observe the anomalous exponent
$\gamma(g^2_*)$ to be in the interval 0.09--0.41, too small
for the desired walking behaviour.   One reason for the
small value is that the fixed point is at relatively small
coupling, $g_*^2 \approx 2$--$3$.  Because $\gamma$ is roughly
linear function of $g^2$, a larger fixed point coupling would
correspondingly increase the anomalous exponent.

\subsection{Particle spectrum in MWTC} 

In principle, the mass spectrum of the theory
should give clear indication of the nature of the
theory at long distances:  if we assume
QCD-like behaviour and chiral symmetry breaking, 
we should
observe massless pseudoscalar Goldstone ``pions,''
whereas other states remain massive as 
as the fermion mass $m_Q \rightarrow 0$.

On the other hand, if the theory has
an infrared fixed point, all states become
massless as $m_Q \rightarrow 0$, as
\begin{equation}
  M \propto m_Q^{1/(1+\gamma(g_*^2))}
\end{equation}
Thus, this gives another way to measure
the mass anomalous exponent.  

\begin{figure}
\centerline{\includegraphics[width=\columnwidth]{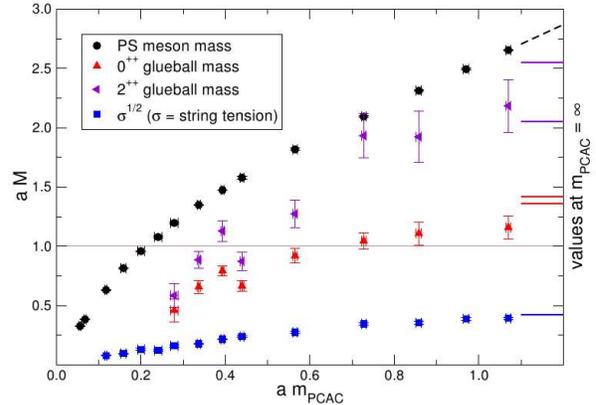}}
\caption{The mass spectrum of MWTC as a function of the
  fermion mass $m_Q a$.  All masses vanish as $m_Q\rightarrow 0$.
  In contrast with QCD, in this case the pseudoscalar is heavier than 
  glueballs or square root of string tension (from \cite{DelDebbio:2009fd}).
}
\label{fig:masses}
\end{figure}

In Figure~\ref{fig:masses} the pseudoscalar meson and glueball
masses and the string tension are shown against the fermion
mass \cite{DelDebbio:2009fd}.  All states become 
massless as the fermion mass vanishes, in accordance with
the IRFP.  In a recent study by Kerrane {\em et al.} \cite{Kerrane:2010xq}
the anomalous exponent $\gamma(g_*^2)$ was measured from the mass spectrum,
with the result that it is compatible with zero and certainly
$\gamma(g_*^2) \ll 1$.

\section{Conclusions}

Large-scale simulations of theories related with technicolor
models started around 3 years ago.  During these years the
activity has strongly increased and there has been 
significant progress in  methodology.  There are now clear indications of the
existence of an infrared fixed point in several candidate theories,
and the approximate location of the conformal window is also roughly known.
Nevertheless, no convincing evidence for a walking couplng 
has been found yet.  However, as the field is still maturing
and the computational effort invested in studies grows,
we can expect definite results in some candidate theories in
the future.

One of the best studied models is the minimal walking technicolor, 
SU(2) gauge field with two adjoint representation fermions.
There are strong indications about the existence of 
an IRFP from the $\beta$-function and excitation spectrum,
at $g^2_{*} \approx 2$--$3$ in SF scheme.  In this 
case the
mass anomaolous exponent $\gamma(g_*^2)$ appears to
be small; too small for the condensate enhancement needed
for walking technicolor.  This seems to be the
case for most candidate models where the anomalous exponent
has been studied.  However, minimal walking technicolor has been
studied only using unimproved Wilson fermions, and
in order to have better control of possible lattice
artifacts calculations with improved actions are needed.

\begin{theacknowledgments}
  This work has been supported by the Academy of Finland grant
  1134018.  The computations described in refs.~\cite{Hietanen:2009az}
  have been done
  at the Finnish IT Center for Science (CSC).
\end{theacknowledgments}

%%%%%%%%%%%%%%%%%%%%%%%%%%%%%%%%%%%%%%%%%%%%%%%%
%% The bibliography can be prepared using the BibTeX program or
%% manually.
%%
%% The code below assumes that BibTeX is used.  If the bibliography is
%% produced without BibTeX comment out the following lines and see the
%% aipguide.pdf for further information.
%%
%% For your convenience a manually coded example is appended
%% after the \end{document}
%%%%%%%%%%%%%%%%%%%%%%%%%%%%%%%%%%%%%%%%%%%%%%%%

%%%%%%%%%%%%%%%%%%%%%%%%%%%%%%%%%%%%%%%%%%%%%%%%
%% You may have to change the BibTeX style below, depending on your
%% setup or preferences.
%%
%%
%% For The AIP proceedings layouts use either
%%%%%%%%%%%%%%%%%%%%%%%%%%%%%%%%%%%%%%%%%%%%

\bibliographystyle{aipproc}   % if natbib is available
%\bibliographystyle{aipprocl} % if natbib is missing

%%%%%%%%%%%%%%%%%%%%%%%%%%%%%%%%%%%%%%%%%%%
%% You probably want to use your own bibtex database here
%%%%%%%%%%%%%%%%%%%%%%%%%%%%%%%%%%%%%%%%%%%
%\bibliography{sample}

%%%%%%%%%%%%%%%%%%%%%%%%%%%%%%%%%%%%%%%%%%%
%% Just a reminder that you may have to run bibtex
%% All of it up to \end{document} can be removed
%% if you don't like the warning.
%%%%%%%%%%%%%%%%%%%%%%%%%%%%%%%%%%%%%%%%%%% 
\IfFileExists{\jobname.bbl}{}
{\typeout{}
  \typeout{******************************************}
  \typeout{** Please run "bibtex \jobname" to optain}
  \typeout{** the bibliography and then re-run LaTeX}
  \typeout{** twice to fix the references!}
  \typeout{******************************************}
  \typeout{}
}

%%%%%%%%%%%%%%%%%%%%%%%%%%%%%%%%%%%%%%%%%%%
%% The following lines show an example how to produce a bibliography
%% without the help of the BibTeX program. This could be used instead
%% of the above.
%%%%%%%%%%%%%%%%%%%%%%%%%%%%%%%%%%%%%%%%%%%

%%
%% End of file `template-8d.tex'.

\end{document}